\documentclass[prl,twocolumn,showpacs]{revtex4}
\usepackage{graphicx}

\begin{document}

\title{Photon Localization in Resonant Media}

\author{A.A.~Chabanov and A.Z.~Genack}

\affiliation{Physics Department, Queens College of CUNY, Flushing, NY 11367}

\date{March, 2001}

\begin{abstract}
We report measurements of microwave transmission over the first five Mie resonances of alumina spheres randomly positioned in a waveguide. Though precipitous drops in transmission and sharp peaks in the photon transit time are found near all resonances, measurements of transmission fluctuations show that localization occurs only in a narrow frequency window above the first resonance. There the drop in the photon density of states is found to be more pronounced than the fall in the photon transit time, leading to a minimum in the Thouless number.
\end{abstract}


\maketitle

Photon localization has been of intense interest since it was first proposed as the analog of electron localization \cite{John}. Accurate measurements of the statistics of the electromagnetic field and intensity and of the total transmission in random samples have deepened our understanding of mesoscopic physics and localization. However, the very possibility of photon localization in three-dimensional systems without substantial crystalline order has been called into question by the barriers to achieving and detecting electromagnetic localization. Unlike electrons that can be trapped by the Coulomb interaction at atomic sites, photons are not bound by individual particles. They are not strongly scattered by particles either, except at Mie resonances where the scattering cross-section can considerably exceed the geometric cross-section. To preserve the strength of scattering, however, particles must be separated by at least the diameter of the scattering cross-section, requiring particle separation that are larger than the wavelength $\lambda$. It would then appear that the Ioffe-Regel-like condition for localization in three dimensions \cite{Bart}, $\ell_{sc}\leq\lambda/2\pi$, where $\ell_{sc}$ is the scattering mean free path, cannot be satisfied by collections of independent resonant scatterers \cite{Ad-review}. Measurements of exponential scaling of transmission \cite{Azi-loc,Ad,Vlas} have not definitively established localization because the possible presence of absorption could also lead to exponential decay of intensity within a sample. However, recent measurements of coherent backscattering in macroporous GaP networks \cite{Schuurmans} along with theoretical predictions \cite{Bart-loc} suggest that the approach to localization can be observed in the rounded backscattering peak from weakly absorbing samples. Also recently, the variance of relative fluctuations has been shown to provide a decisive test for localization, even in the presence of strong absorption \cite{Nature}. This provides a sure guide in the search for localization and raises the possibility that measurements of key scattering parameters can be used to sort out the precise material and structural characteristics that may edge samples towards and potentially across the localization threshold. 

In this Letter, we report microwave measurements of the frequency variation of three key localization parameters in quasi-1D samples composed of randomly positioned alumina spheres in an open copper tube. These are the Thouless number \cite{Thouless}, which is the ratio of the width to the spacing between quasi-states of a random medium, $\delta=\delta\nu/\Delta\nu$, the dimensionless conductance $g$ \cite{Gang}, and a parameter $g^{\prime}$ \cite{Nature}, which represents the inverse of the variance of transmission normalized to its ensemble average value. These parameters capture, respectively, the relation to localization of average dynamics, average static transmission, and static fluctuations. Their measurement makes it possible to identify a narrow window of localization above the first Mie resonance and to probe a constellation of factors that foster localization, including size, concentration, and structural correlation.

We report the first direct measurement of the Thouless number. The Thouless criterion identifies the onset of localization as the point at which the level width $\delta\nu$ becomes smaller than the level spacing $\Delta\nu$ \cite{Thouless}. Beyond this point, transport is inhibited since modes in different blocks of a sample do not overlap. The level width $\delta\nu$ is identified with the field correlation frequency \cite{Azi-EuL}. This is inversely related to the width of the time-of-flight distribution of the transmitted wave and to the average transit time through the medium, $\tau$ \cite{Patrick1}. The level spacing $\Delta\nu$ is the inverse of the density of states, $\Delta\nu=1/N(\nu)$, thus giving $\delta=\delta\nu N(\nu)$. The hurdles to localizing radiation, even in resonant media, can be seen from a dynamical perspective. The precipitous drop in $\delta\nu$ expected at resonance, reflecting the sharp peak in $\tau$, does not necessarily result in a diminished value of $\delta$ because it is countered by an increasing value of $N(\nu)$.

The dimensionless conductance $g$ is given by the sum, $g=\Sigma_{ab}T_{ab}$ \cite{Fisher}, where $T_{ab}$ is the transmission coefficient between incident mode $a$ and output mode $b$. In electronic systems, $g$ can be obtained from the measurement of the conductance, $G=(e^{2}/h)g$, whereas for classical waves it can be found from the measurement of the total transmission $T_{a}=\Sigma_{b}T_{ab}$, since $g=N\langle T_{a}\rangle$, where $N$ is the number of transverse modes given by $N=Ak^{2}/2\pi$, where $A$ is the area of the cross-section of a sample and $k$ is the wave number, and the angle brackets denote the ensemble average. For diffusive waves in nonabsorbing samples of length $L$, $g=N\ell/L$, where $\ell$ is the transport mean free path. The parameter $g$ is essentially the number of diffusing channels in a sample and so the localization threshold is reached when its value falls below unity \cite{Imry}. This is consistent with the Thouless criterion for localization, since for diffusive waves, $g=\delta>1$ \cite{Gang}.

In the diffusive limit and in the absence of absorption, the variance of the total transmission normalized to its ensemble average is given by $var(s_{a})=2/3g$, where $s_{a}=T_{a}/\langle T_{a}\rangle$ \cite{Kogan,Rossum,Marin}. This allows us to construct a measure of fluctuations, $g^{\prime}=2/3var(s_a)$, which reduces to $g$ in the above limit. The localization transition should occur when $g^{\prime}\simeq 1$ \cite{Nature}. Since $var(s_a)$ and $var(s_{ab})$, where $s_{ab}=T_{ab}/\langle T_{ab}\rangle$, are related by $var(s_{a})={1\over 2}(var(s_{ab})-1)$ \cite{Nature,Kogan}, we are able to evaluate $g^{\prime}$ either from measurements of total transmission, using an integrating sphere \cite{Marin}, or from local measurements of intensity \cite{Nature}. 

Measurements of field transmission spectra are carried out in low-density collections of 0.95-cm-diameter alumina spheres (99.95\% Al$_2$O$_3$) with use of a Hewlett-Packard 8772C network vector analyzer. Low densities are produced by embedding the alumina spheres in Styrofoam spheres of low refractive index. The index of refraction $n$ of the alumina spheres is inferred from a comparison of the measured extinction coefficient of the coherent intensity $I_{c}$ in 3-D samples with an alumina volume fraction $f = 0.010$ with the calculated scattering cross-section $\sigma_{sc}$ for isolated alumina spheres. In the independent-scattering approximation, $\ln I_{c}=-n_{sc}\sigma_{sc}L$, where $n_{sc}$ is the number density of the alumina scatterers. The natural logarithm of $I_{c}$ is compared to calculations of $\sigma_{sc}$ with $n = 3.14$ in Fig.~1. This gives the closest correspondence between the positions of the Mie resonances and those of the dips in $\ln I_{c}$.

\begin{figure}[b!]
\includegraphics[width=\columnwidth]{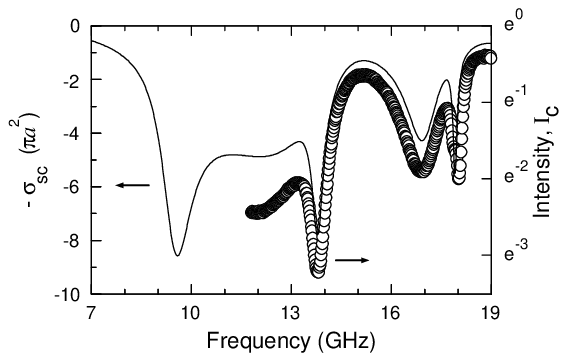}
\caption{Comparison of the scattering cross-section $\sigma_{sc}$ of a 0.95-cm-diameter alumina sphere (solid) and the extinction of the coherent intensity $I_{c}$ in a 3-D alumina sample with $f=0.010$ (circles). In calculations, the alumina sphere is assumed to be lossless and the refractive index is adjusted so that the positions of the Mie resonances are located at the dips in the experimental data. The curve shown is for $n=3.14$.}
\end{figure}

\begin{figure}[b!]
\includegraphics[width=\columnwidth]{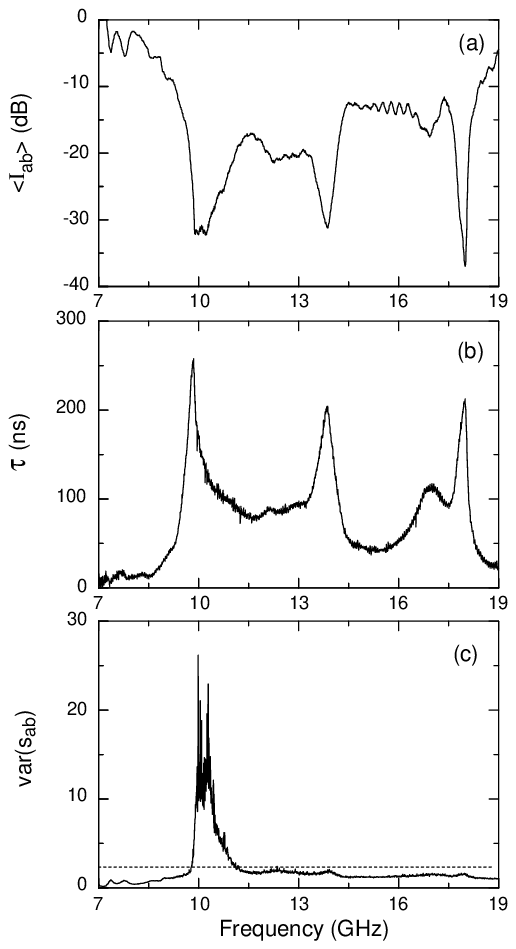}
\caption{(a) Average transmitted intensity $\langle I_{ab}\rangle$, (b) average photon dwell time $\tau$, and (c) $var(s_{ab})$ in a quasi-1D alumina sample of $L=80$ cm and $f=0.068$. The dashed line is the localization threshold.}
\end{figure}

Measurements of microwave propagation and localization are made in quasi-1D samples with an alumina volume fraction of $f=0.068$ contained in a 7.3-cm-diameter copper tube of $L=80$ cm and $L=12$ cm.  The alumina spheres are put at centers of 1.9-cm-diameter Styrofoam spheres. An ensemble of 5,000 statistically equivalent random samples is created by rotating the tube between successive field measurements. The incident mode $a$ is parallel to the sample axis. The outgoing field is detected at a point on the output surface of the sample. For each sample configuration, the field spectrum yields the corresponding frequency variation of the transmitted intensity $I_{ab}$. The ensemble average $\langle I_{ab}\rangle$ for $L=80$ cm samples, shown in Fig.~2a, exhibits distinct dips near each of the Mie resonances. The resonant character of scattering is further indicated by the sharp peaks in the spectrum of the average photon dwell time, seen in Fig.~2b. The transit time is given by $\tau=\langle s_{ab}d\phi_{ab}/ d\omega\rangle$, where $\phi_{ab}$ is the phase of the field and $\omega$ is the angular frequency \cite{Patrick2}. Since low transmission can be due to absorption and long dwell time can be associated with microstructure resonances, these indicators do not provide a definitive measure of the closeness to the localization threshold. This can be obtained, however, from the measurement of $var(s_{ab})$, shown in Fig.~2c. For diffusive waves obeying Rayleigh statistics, $var(s_{ab})=1$. Lower values below 8.5 GHz indicate a significant ballistic component in the transmitted field, while higher values indicate the presence of substantial long-range correlation. The horizontal dashed line in Fig.~2c represents the localization threshold, $var(s_{ab})=7/3$, which corresponds to $var(s_{a})=2/3$ or, equivalently, $g^{\prime} = 1$. The localization threshold in this quasi-1D sample is crossed in a narrow frequency range above the first Mie resonance. Large fluctuations in measurements of $var(s_{ab})$ above the localization threshold, seen in Fig.~2c, are a result of the greatly enhanced tail of the intensity distribution for localized radiation \cite{Nature}.
 
In order to take a closer look at the localization region, the data in Fig.~2 are superimposed and displayed using an expanded scale in Fig.~3a. The shift of extrema in $\langle I_{ab}\rangle$ and in $var(s_{ab})$ to frequencies above the maximum in $\tau$ indicates that localization occurs above the resonant peak. Since localization occurs within a trough of $\delta$ in a frequency range in which $\tau$ is smaller than its peak value by a factor of three, $N(\nu)$ must drop from its peak value on resonance by an even greater factor. We also expect that in the absence of absorption, $g^{\prime}\simeq g$, and that the peak in $var(s_{ab})$ should occur at the same frequency as the dip in $\langle I_{ab}\rangle$ rather than at higher frequency as seen in Fig.~3a. The shift observed in the corresponding spectra in Fig.~3a is principally a result of the peak in photon dwell time, which increases absorption and suppresses $\langle I_{ab}\rangle$ more than $var(s_{ab})$ at the resonance. Enhanced absorption on resonance is also the cause of the drops in transmission at higher Mie resonances in Fig.~2a.

\begin{figure}[t!]
\includegraphics[width=\columnwidth]{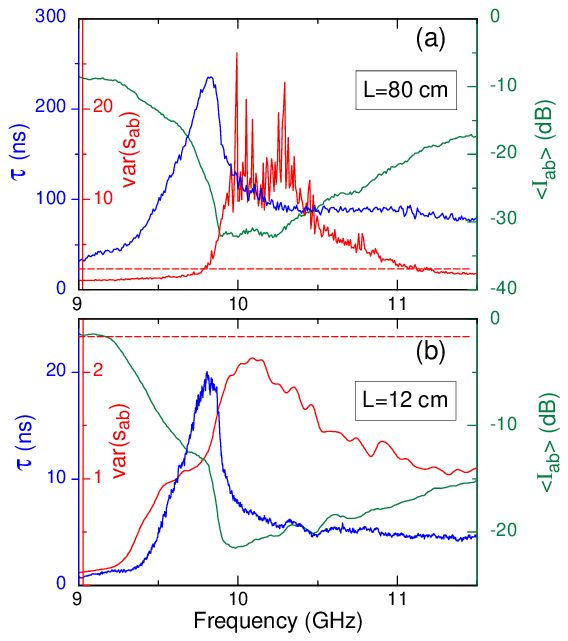}
\caption{(color) $\langle I_{ab}\rangle$, $\tau$ and $var(s_{ab})$ superimposed over the frequency range of the first Mie resonance: (a) $L=80$ cm, (b) $L=12$ cm. The dashed line is the localization threshold.}
\end{figure}

\begin{figure}[t!]
\includegraphics[width=\columnwidth]{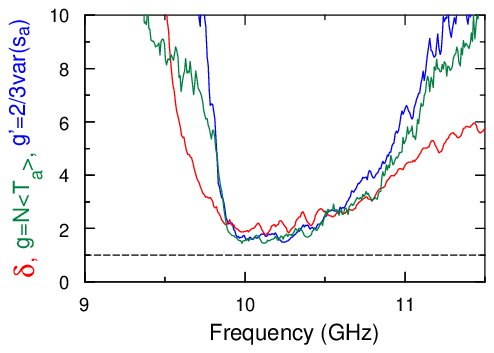}
\caption{(color) Localization parameters $\delta$, $g$ and $g^{\prime}$ over the frequency range of the first Mie resonance in an alumina sample of $L=12$ cm. The dashed line is the localization threshold.}
\end{figure}

To investigate the factors that incline a sample towards localization, we carry out the measurements presented in Fig.~3a in a shorter sample of $L=12$ cm (Fig.~3b), so that $\ell<L<\xi, L_{a}$, where $\xi = N\ell$ is the localization length and $ L_{a}$ is the exponential absorption length. In this weakly absorbing sample, we find that transport is diffusive above 9.6 GHz, where $1<var(s_{ab})<7/3$  (Fig.~3b), so that $g$, $g^{\prime}$, and $\delta$ can be associated unambiguously with the proximity to the localization threshold. The quantities $g$ and $g^{\prime}$ are obtained from measurements of the total transmission $T_{a}$, via expressions $g=N\langle T_{a}\rangle$ and $g^{\prime}=2/3var(s_{a})$. Good agreement between $g$ and $g^\prime$ is found, as seen in Fig.~4. Also notice that, since the minimum of $g$ occurs above the resonance, the minimum of $\ell$ must also lie above that of $\ell_{sc}$. 

To examine more closely the balance of competing effects near the resonance, we measure $\delta\nu$, $\Delta\nu$, and $\delta$. The level width $\delta\nu$ is determined from measurements of the field correlation function with frequency shift. We find that $\delta\nu$ shown in Fig.~5b is related to the average diffusion time (Fig.~3b), $\tau=L^{2}/6D=1/6\delta\nu$. The level spacing $\Delta\nu$ is found by measuring transmission between two probes in the tube with copper end caps. In this case, line widths are narrowed and mode spacing can be measured (Fig.~5a). When determining $\Delta\nu$ both positive and negative peaks in spectra of $d\phi_{ab}/d\omega$ are accounted. The negative peaks appear because of interference between two closely spaced modes. A comparison of $\Delta\nu$ calculated for an ensemble of 360 sample configurations and $\delta\nu$ is shown in Fig.~5b. We find that, whereas $\delta\nu$ and  $\Delta\nu$ attain their minimum values at the resonance, their ratio is a minimum above the resonance. In the trough of $\delta$, where the measurements of $\Delta\nu$ are most reliable by virtue of the lowest values of $N(\nu)$, $\delta$ is in a good agreement with $g$ and $g^{\prime}$ (Fig.~4).

\begin{figure}
\includegraphics[width=\columnwidth]{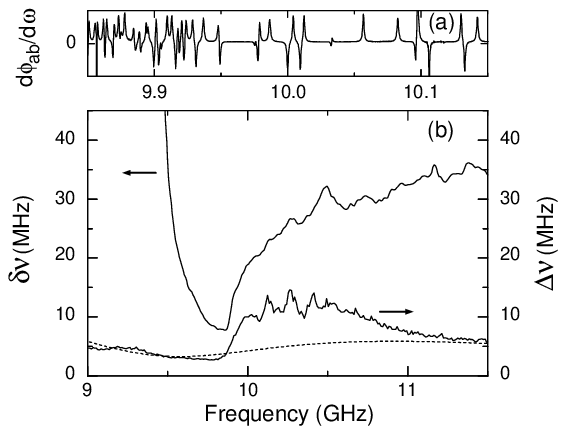}
\caption{(a) Part of a spectrum of $d\phi_{ab}/d\omega$ for a single configuration with the peaks marking the mode frequencies and (b) the level width $\delta\nu$ and the average level spacing $\Delta\nu$ in a $L=12$ cm alumina sample. At all frequencies, $\delta\nu>\Delta\nu$. The dashed line is the inverse of $N(\nu)=(8\pi\nu^{2}/v_{0}^{3})(v_{p}/v_{E})V$ \cite{Bart}, where $v_{p}$ and $v_{E}$ are the phase \cite{Bart} and the energy \cite{Bart-vE} velocity, respectively, $v_{0}$ is speed of light in the effective medium, and $V$ is the sample volume.}
\end{figure}

The increase above resonance in $\Delta\nu$, obtained from a calculation of $N(\nu)$ in the independent-scattering approximation \cite{Bart}, is well below the measured increase (Fig.~5b). This suggests that increasing the density of resonant scatterers beyond the point at which they begin to interact suppresses the density of states and thereby fosters localization. A decrease in $N(\nu)$, on the other hand, was predicted by John \cite{John-PBG} for nearly periodic samples. Indeed, for periodic samples exhibiting a complete photonic band gap, the vanishing of $N(\nu)$ automatically results in localization within a pseudogap, when disorder is introduced. We find that, when the alumina spheres are displaced by half of their radius from the centers of the embedding Styrofoam spheres, $var(s_{ab})$ is only slightly suppressed. For a $L=55$ cm alumina sample, $var(s_{ab})$ drops from 6.0 to 4.5 at its maximum. This small reduction, for a considerable broadening of the radial correlation function, suggests that collective scattering from randomly positioned alumina spheres itself and not only their residual positional order is the chief reason for the sharp drop in $N(\nu)$ above the first Mie resonance.

In conclusion, we have measured the dimensionless conductance, the variance of transmission fluctuations, and the Thouless number in low-density collections of alumina spheres in a quasi-1D geometry. We find these are equivalent measures of localization for $L<L_a, \xi$.  For samples longer than $L_{a}$, only $var(s_a)$ serves as a reliable guide to localization, which is found in a narrow frequency range above the first Mie resonance in the longer sample. Localization does not occur on resonance because the narrowing of the level width is offset by the enhancement of the density of states. However, a sharp drop in $N(\nu)$ above resonance leads to a minimum in $\delta$. The drop in $N(\nu)$ and the consequent appearance of localization are the result of collective scattering. 

\begin{acknowledgments}
We are grateful to A. Geyfman, E. Kuhner and Z. Ozimkowski for their help in constructing the experimental apparatus. We thank V. Kopp and O. Pudeyev for assistance in analysis of $\Delta\nu$ and B.A. van Tiggelen for valuable discussions. This work was supported by the National Science Foundation and U.S. Army Research Office.
\end{acknowledgments}

\end{document}